\documentclass[conference]{IEEEtran}
\IEEEoverridecommandlockouts
\usepackage{cite}
\usepackage[numbers,sort&compress]{natbib}
\usepackage{amsmath,amssymb,amsfonts}
\usepackage{algorithmic}
\usepackage{graphicx}
\usepackage{booktabs}
\usepackage{textcomp}
\usepackage{xcolor,soul}
\usepackage{subfig}
\usepackage[pagebackref=true,breaklinks=true,letterpaper=true,colorlinks,bookmarks=false]{hyperref}
\usepackage{fancyhdr}
\def\BibTeX{{\rm B\kern-.05em{\sc i\kern-.025em b}\kern-.08em
    T\kern-.1667em\lower.7ex\hbox{E}\kern-.125emX}}
    
\makeatletter
\def\footnoterule{\relax%
  \kern-5pt
  \hbox to \columnwidth{\hfill\vrule width 1\columnwidth height 0.4pt\hfill}
  \kern4.6pt}
\makeatother

\newcommand{\dafinal}[1]{\sethlcolor{green}[#1]}     

\begin{document}
\title
{
    Caffe \textit{Barista}: \\ Brewing Caffe with FPGAs in the Training Loop
}


\author{\IEEEauthorblockN{Diederik Adriaan Vink\IEEEauthorrefmark{2}\IEEEauthorrefmark{1}, 
    Aditya Rajagopal\IEEEauthorrefmark{2}\IEEEauthorrefmark{1}, 
    Stylianos I. Venieris\IEEEauthorrefmark{3},
    Christos-Savvas Bouganis\IEEEauthorrefmark{2}}\\
    \IEEEauthorblockA{\IEEEauthorrefmark{2}\textit{Intelligent Digital Systems Lab, Imperial College London\hspace{+0.75cm}\IEEEauthorrefmark{3}\textit{Samsung AI Center, Cambridge}}\\ 
    \IEEEauthorblockA{\{diederik.vink14, aditya.rajagopal14, ccb98\}@ic.ac.uk\hspace{+2cm}s.venieris@samsung.com\\\\
    \small\textit{{\IEEEauthorrefmark{1}Indicates equal contribution.}}}}
}

\maketitle

\thispagestyle{fancy}
\chead{PREPRINT: Accepted at the 30th International Conference on Field Programmable Logic and Applications (FPL), 2020}
\renewcommand{\headrulewidth}{0pt}

\begin{abstract}
As the complexity of deep learning (DL) models increases, their compute requirements increase accordingly. 
Deploying a Convolutional Neural Network (CNN) involves two phases: training and inference. 
With the inference task typically taking place on resource-constrained devices, a lot of research has explored the field of low-power inference on custom hardware accelerators. 
On the other hand, training is both more compute- and memory-intensive and is primarily performed on power-hungry GPUs in large-scale data centres.  
CNN training on FPGAs is a nascent field of research. 
This is primarily due to the lack of tools to easily prototype and deploy various hardware and/or algorithmic techniques for power-efficient CNN training. 
This work presents \textit{Barista}, an automated toolflow that provides seamless integration of FPGAs into the training of CNNs within the popular deep learning framework Caffe. 
To the best of our knowledge, this is the only tool that allows for such versatile and rapid deployment of hardware and algorithms for the FPGA-based training of CNNs, providing the necessary infrastructure for further research and development.
\end{abstract}


\section{Introduction}
Convolutional Neural Networks (CNNs) are one of the primary components across a wide variety of AI tasks, from face recognition~\cite{face_recognition_sota} to drone navigation~\cite{kouris_drone2018iros}. 
The process of deploying a CNN involves two stages. 
First, the CNN is trained~\cite{krizhevsky_imagenet_2012} on a large amount of labelled data from a task-specific dataset. 
The second stage involves performing inference on unseen inputs for either classification~\cite{wang_cnn-rnn_2016, hershey_cnn_2017}, detection~\cite{girshick_fast_2015, kouris_informed_2019} or segmentation~\cite{long_segmentation}. 
Inference is usually performed in resource- and power-constrained environments and hence the CNN needs to be deployed on power-efficient embedded devices such as mobile System-on-Chips (SoCs)~\cite{embench_2019}, or FPGA-based platforms~\cite{crockett_zynq_2014}. 
To this end, significant efforts have been invested towards custom FPGA-based accelerator designs for the inference stage of CNNs~\cite{venieris_toolflows_2018, dicecco_caffeinated_2016, Guan-2017, jiao_accelerating_2017, light_opu_2020}.

Due to its large computational demands and the massive datasets, CNN training is usually performed on powerful GPUs hosted in private clusters or data centres. 
For such setups, the power and cooling infrastructure constitutes the dominant factor of the operational expenses~\cite{kozyrakis_2013}. 
With GPUs being power-hungry, they become costly platforms to maintain. 
This fact has led industrial players to equip their servers with custom ASICs, such as Google TPU~\cite{Google-TPU}, Graphcore IPU~\cite{Graphcore-IPU} and Amazon Inferentia~\cite{amazon_inferentia}. 
Nevertheless, the long development time and time-to-market together with their fixed functionality limit the ASICs' ability to exploit model-specific optimisations and support the latest fast-paced algorithmic advances.

In this context, FPGAs constitute a promising alternative \cite{2018_cascade_cnn, 2018_f-cnnx, 2018_fpga_convnet}. 
Due to their customisability and reconfigurability, FPGAs can attain competitive performance and power efficiency for flexible, power- and cost-efficient development and deployment of DL training workloads.
At the same time, public cloud providers are increasingly offering access to FPGA platforms~\cite{f1_aws,MSR-Brainwave,huawei_fpga_cloud}, increasing their accessibility and making the rapid low-cost deployment of FPGA designs feasible.
Nevertheless, so far, FPGA-based CNN training has only slightly been explored~\cite{wenlai_zhao_f-cnn_2016, Geng-2019, gcn_train_fpga_2020} largely due to the lack of tools to easily prototype and deploy various hardware and/or algorithmic techniques for efficient CNN training. 


The primary contribution of this work is \textit{Barista}, an open-source toolchain\footnote{\href{https://github.com/ICIdsl/caffe\_fpga.git}{https://github.com/ICIdsl/caffe\_fpga.git}} integrated into the widely used DL framework Caffe~\cite{caffe_2014}, that enables the rapid prototyping and deployment of FPGA-based kernels for CNN training. 
Additionally, the work provides a memory-aware model for the execution of an FPGA-based general matrix multiply (GEMM) kernel along with an initial HLS implementation of this kernel.
In this manner, \textit{Barista} allows both hardware researchers and machine learning experts to explore novel hardware and algorithmic techniques respectively for power-efficient training. 

\section{Background \& Related Work} \label{sec:background}


\textit{Barista} enables rapid prototyping and deployment of hardware accelerators for DL training. 
Here, key challenges and requirements of architectures targeting such workloads are described and existing work on FPGA-based training is reviewed.

\textbf{DNN Training.}
DNNs are generally trained offline, following an iterative process~\cite{sgd_variants_2010}. 
Each iteration comprises three steps: a \textit{forward pass}, a \textit{backward pass} (through backpropagation) and a \textit{weight update}. 
The forward pass (inference task) calculates the loss for a given input. 
The backward pass employs the backpropagation algorithm to compute the gradient of the loss with respect to the trainable weights; the weight update step updates the weights using these gradients. 
In each iteration, training operates over mini-batches of labelled inputs from the training set; in this respect, \textit{throughput} is the primary performance metric of interest, in contrast to the inference task's low-latency requirements. 
Furthermore, with power and cooling being a critical expense in both public and private clusters, power efficiency constitutes another decisive metric.

\textbf{FPGA-based CNN Training.}
Existing work on FPGA-based CNN training can be taxonomised in two broad categories: 1)~costly multi-FPGA systems~\cite{wenlai_zhao_f-cnn_2016, Geng-2019} and 2)~highly customised accelerators~\cite{Luo-2019, FOX-2019}.
F-CNN~\cite{wenlai_zhao_f-cnn_2016} enabled run-time reconfiguration through overlapping computation by utilising multi-device platforms.
FPDeep~\cite{Geng-2019} proposed a load balancing scheme to train CNNs across multiple FPGAs ($>$15). 
Focusing on single-FPGA setups, DarkFPGA~\cite{Luo-2019} employed a batch-oriented data layout scheme optimised for a specific hardware design and is not applicable to other accelerators. 
Furthermore, \cite{FOX-2019} designed a low-precision training accelerator through hardware-algorithm co-design. 
By replacing the GEMM, the approach of this work enables training of any DNN that uses matrix multiplication.

All aforementioned works adopt proprietary front-ends, lacking integration with the traditional machine learning frameworks and the support that they provide, and/or propose workload-specific architectures that cannot be used for general DNN training.
Similar to this work, \textit{FeCaffe} \cite{He-2020} proposes a system that integrates Caffe with an FPGA. 
However, \textit{Barista} provides more details on the challenges of developing a custom accelerator, as well as provides an analytical model for performance prediction allowing the tuning of the framework under diverse workloads. 
Additionally, unlike \textit{FeCaffe}, the Caffe integration of the proposed system will be open-sourced to promote adoption and research, and can be deployed on any AWS F1 instance using the GEMM bitstream provided.

\section{System Design} \label{sec:system_design}

The \textit{Barista} tooflow integrates with the Caffe framework and targets systems with PCIe-based FPGA accelerators. 
To this end, \textit{Barista} consists of three components: 1)~a software integration layer that enables the seamless integration of the FPGA accelerator with Caffe, 2)~an FPGA-based hardware accelerator, and 3)~an OpenCL runtime that orchestrates the CNN execution between a host CPU and the FPGA. 
Upon deployment, the FPGA device runs a kernel that is responsible for executing the matrix multiplications involved in the forward and backward pass of the CONV layers throughout CNN training. 
The CPU executes all other operations of the training process and coordinates the offloading of computations to the FPGA.

    \subsection{Caffe GEMM Execution Flow} \label{sec:caffe_flow}
    This section provides a description of Caffe's native execution flow for CONV layers.
    Initially, Caffe selects which platform (CPU/GPU/FPGA) to execute on.
    Next, for each batch in each layer Caffe calls the batch-level GEMM function.
    At this point, the implementations for forward and backward pass start to deviate.
    For the forward pass, Caffe calls \texttt{im2col} on all the inputs and weights to convert them to matrices in order to execute convolutions as GEMMs. 
    For the backward pass, the gradients w.r.t the weights are calculated by multiplying the inputs with the gradients w.r.t the output.
    Then, it calculates the gradient w.r.t to the input for each element in the batch by multiplying the weights with the gradient w.r.t the output.
    All these matrices are split into tiles (Section \ref{sec:accel_arch}) and then fed to the FPGA GEMM kernel.
    As the forward pass is a GEMM, \texttt{im2col} is not required for backpropagation.

\subsection{Accelerator Architecture for CNN Training}
\label{sec:accel_arch}

The developed hardware architecture is designed to meet the typical requirements of CNN training workloads (Section~\ref{sec:background}). 
The core compute block comprises a parametrised systolic array for the execution of blocked GEMM that is reused across both the forward and backward passes of CONV layers. 

\begin{figure}[t]
	\centering
	\includegraphics[trim={3cm 5.5cm 2cm 5cm},clip,width=1\linewidth]{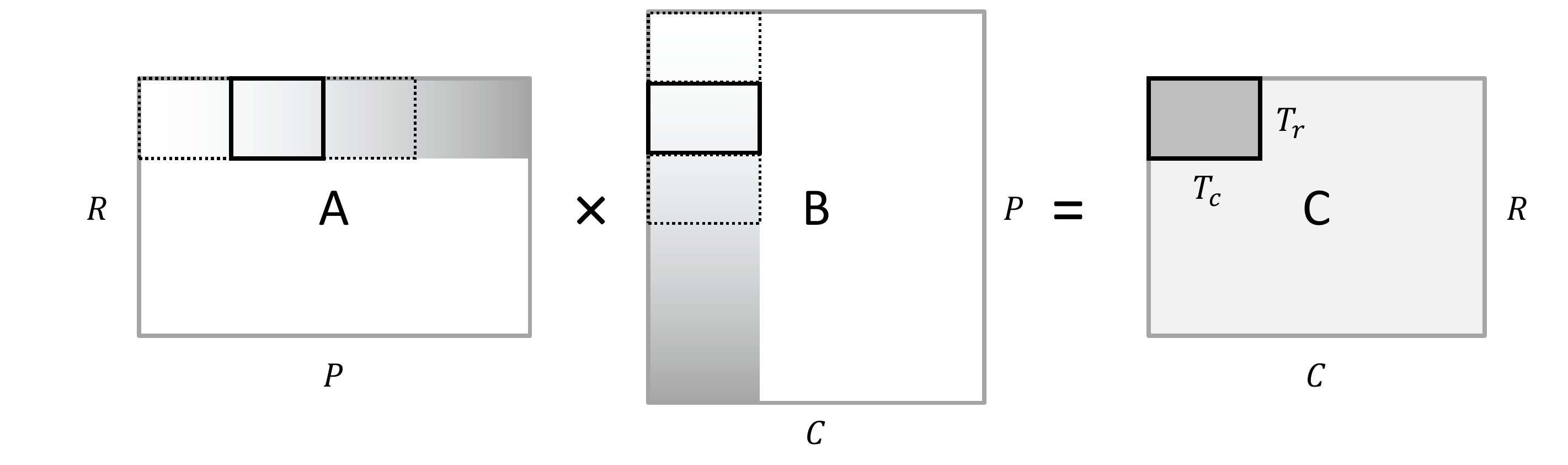}
	\caption{Overview of the adopted blocked GEMM strategy.}
	\label{fig:blocked_gemm}
\end{figure}

\textbf{Blocked GEMM:} 
As shown in Fig.~\ref{fig:blocked_gemm}, in the operation ($\textbf{C}$$=$$\textbf{AB}$), matrices $\textbf{A}$, $\textbf{B}$ and $\textbf{C}$ are partitioned into tiles. 
Matrix $\textbf{A}$ is partitioned into $\left\lceil \frac{R}{T_r} \right\rceil \cdot \left\lceil \frac{P}{T_p} \right\rceil$ tiles of size $T_r \times T_p$, matrix $\textbf{B}$ is partitioned into $\left\lceil \frac{C}{T_c} \right\rceil \cdot \left\lceil \frac{P}{T_p} \right\rceil$ tiles of size $T_p \times T_c$ and the output matrix $\textbf{C}$ is partitioned into $\left\lceil \frac{R}{T_r} \right\rceil \cdot \left\lceil \frac{C}{T_c} \right\rceil$ tiles of size $T_r \times T_c$. 
If $\frac{R}{T_r} \notin \mathbb{Z}$, then zeros are added to dimension $R$ until $\frac{R}{T_r} \in \mathbb{Z}$.
The same applies to $P$ and $C$ and this process will be referred to as \textit{Tiling} through the rest of the paper.

%

From an operational perspective, the accelerator computes one tile of the output matrix $\textbf{C}$ at a time, until all output tiles are computed. 
For the computation of a single tile, $\left\lceil \frac{P}{T_p} \right\rceil$ tiles from matrix $\textbf{A}$ and $\textbf{B}$ are processed. 
In the implemented blocking strategy, each output tile is cached in the on-chip memory of the accelerator until it has been fully formed. 
Consequently, the intermediate results of the tile are reused $\left\lceil \frac{P}{T_p} \right\rceil$ times before they are written back to the external memory, relaxing the bandwidth requirements of the accelerator.

\begin{figure}[t]
	\centering
	\includegraphics[trim={0cm 0cm 0cm 0cm},clip,width=1\columnwidth]{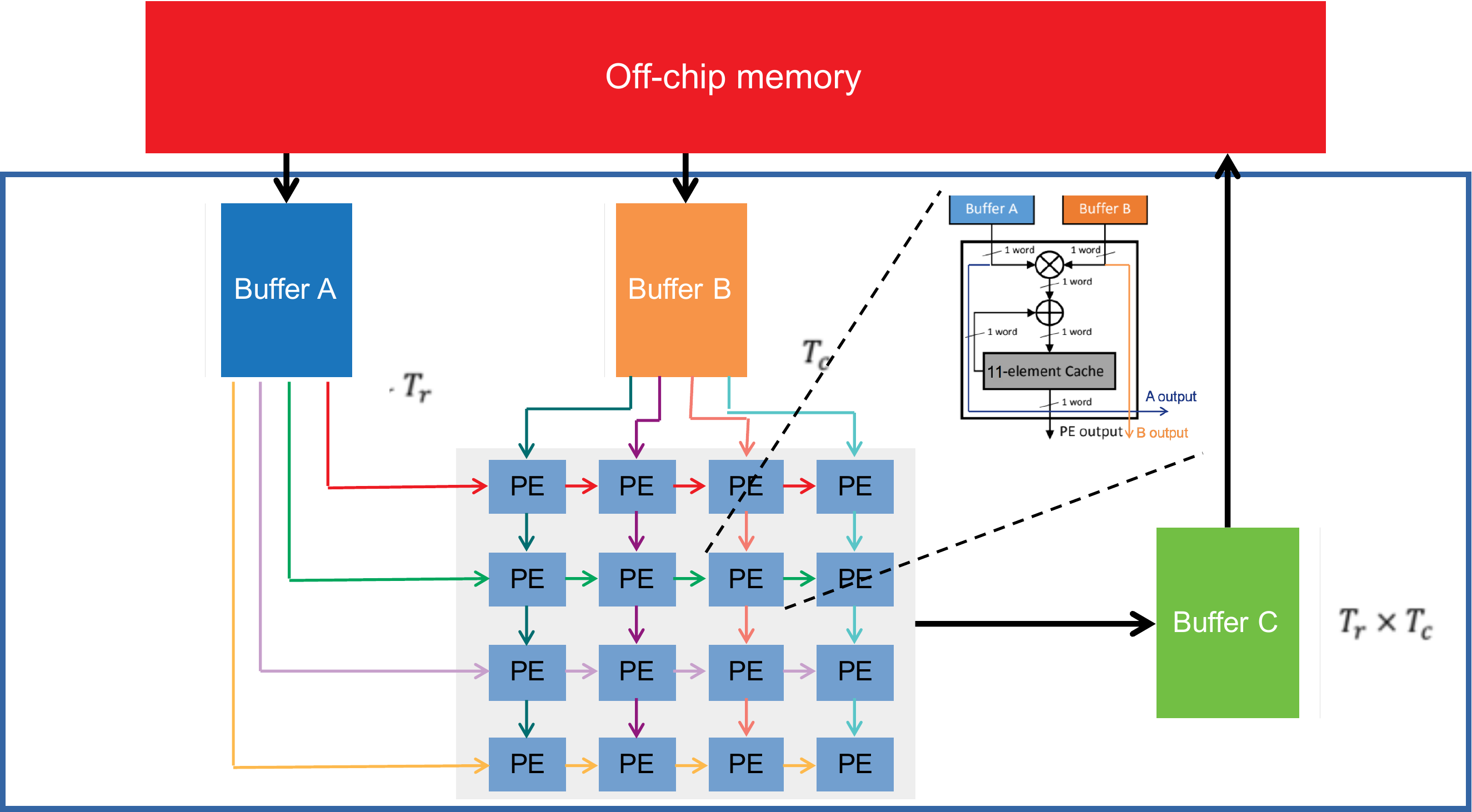}
	\caption{Diagram of the systolic-array design.}
	\label{fig:systolic_array}
\end{figure}


\textbf{Hardware Architecture:} 
Fig.~\ref{fig:systolic_array} shows the adopted hardware architecture for accelerating the blocked GEMM algorithm.
The core of the design is a mesh of processing elements (PEs) and has a throughput of one output per cycle, when the necessary data are available in buffer A and B.
The dimensions of the mesh are compile-time configurable with a total of $T_r \times T_c$ PEs. 
All of the inputs to the GEMM are pre-processed by the CPU into a tiled layout that is sequentially stored in memory.
To better utilize the memory bandwidth, the two matrices to be multiplied are sent to the off-chip memory on the FPGA board in one transaction.
When the execution of the kernel is triggered, the input tiles $T_r \times T_p$ and $T_p \times T_c$ are burst-read from the off-chip memory into buffers A and B which are stored on on-chip memories (\textit{i.e.} BRAMs). 



\textbf{Processing Element:} 
Fig.~\ref{fig:systolic_array} also shows the internal design of each PE. 
Each PE is responsible for computing one element of a tile of the output matrix. 
From a hardware perspective, each PE contains a single multiply-accumulate unit and a local cache for storing the intermediate results of the output, until the final result is ready and written out to external memory. 
The dataflow depicted in Fig.~\ref{fig:systolic_array} enables efficient data passing between PEs in a pipelined fashion, saving routing resources and improving the scalability potential of the design.

\textbf{Precision-aware interleaving:}
\label{sec:interleaving}
Depending on the adopted precision and target device, the latency of a multiplier is $Q$ cycles.
As a result, a direct implementation would require each input to wait for $Q$ cycles until the previous result would be ready for the accumulation.  
To alleviate this, when $Q$$>$$1$, we employ an interleaving technique that computes $Q$$+$$1$ independent intermediate results in a pipelined manner, storing them in the PE's cache.
As a final step, all $Q$$+$$1$ partial values are accumulated into the final result. 
From a performance perspective, this strategy enables a throughput of 1. 

\subsection{OpenCL Runtime} \label{sec:fpga_kernel}
The CPU-FPGA interactions and the FPGA execution are orchestrated by \textit{Barista}'s OpenCL runtime. 
Prior to performing a GEMM operation, this module is responsible for allocating the necessary memory and tiling the input matrices given the selected tile sizes.
Next, it coordinates all CPU-FPGA transfers, launches the FPGA execution and finally collects and untiles the final result to comply with the expected GEMM output format. 
The runtime is executed by the host CPU and employs \texttt{aligned\_storage} vectors for tiles to ensure FPGA word aligned storage on the off-chip memory.


\section{Performance Model} \label{sec:perf_model}
To select the dimensions of the mesh that would yield the highest performance, a performance model was built which estimates the attainable execution time of the hardware design. 
The performance model consists of two components: 1)~the estimated execution time for the processing of a matrix multiplication by the systolic array (Eq.~(\ref{equ:mem_model_latency})) and 2)~the estimated memory transfer time for transferring the matrices $\textbf{A}$, $\textbf{B}$ and $\textbf{C}$ between the host and the FPGA's off-chip memory.

\textbf{Off-chip memory transfer time:}
The design requires $T_r + T_c$ inputs per cycle. 
By denoting the wordlength (bits) of an element by $WL$
the required data that needs to be accessed from the off-chip memory is $\text{Data}_{\text{mem}}$, where the last term captures the data written back to the memory per tile:
\begin{equation*}
\text{Data}_{\text{mem}} =  WL \cdot \left\lceil \frac{R}{T_{r}} \right\rceil 
    \cdot \left\lceil \frac{C}{T_{c}} \right\rceil 
    \cdot \left( \left( T_r \cdot P + T_c \cdot P \right) + T_c \cdot T_r  \right) 
\label{equ:mem_model_data}
\end{equation*}
Overall, given a memory bandwidth of $B_{\text{mem}}$, the latency for accessing the off-chip memory per matrix multiplication is:

\begin{equation}
\text{Latency}_{\text{mem}} = \frac{\text{Data}_{\text{mem}}}{B_{\text{mem}}}
\label{equ:mem_model_latency}
\end{equation}

\textbf{Compute time:}
The number of clock cycles that is needed for the developed system to process the matrix-multiplication computation (\textit{i.e.} \textbf{C=AB}, \textbf{A} is a $R \times P$ matrix, and \textbf{B} is a~$P \times C$ matrix and the output tile size is $T_r \times T_c$) is:
\begin{equation}
    \begin{aligned}
    \text{Cycles}_{\text{compute}}& =
        \left\lceil \frac{R}{T_{r}} \right\rceil 
        \cdot \left\lceil \frac{C}{T_{c}} \right\rceil
        \cdot \\ 
        & \left(
            \left(
                \left\lceil \frac{P}{T_{p}} \right\rceil
                \cdot 
                \left(
                    T_{p} + T_{c} + T_{r} - 2
                \right) 
            \right)
            + (Q+1)^{2}  
        \right)
    \end{aligned}
\label{equ:exec_time}
\end{equation}

\textbf{IP execution time:}
Total GEMM kernel execution latency, when the data are already available in the off-chip \mbox{memory, is:}

\begin{equation}
\text{Latency}_\text{total} =  
\frac{\text{Cycles}_\text{compute}}{f_{\text{clk}}} + 
\text{Latency}_{\text{mem}} 
\label{equ:mem_model_latency}
\end{equation}
where $f_{\text{clk}}$ denotes the clock frequency of the FPGA device.

\textbf{PCIe transfer time:}
The PCIe transfer time captures the latency for the communication of the data from the CPU to off-chip memory. 
$\text{Data}_{\text{PCIe}} = {WL} \cdot \left( R \cdot P + C \cdot P + R \cdot C \right)$ captures the data to be transferred. 
Eq.~(\ref{equ:pci_model_latency}) captures the transfer latency given the PCIe bandwidth $B_{\text{PCIe}}$:
\begin{align}
\text{Latency}_{\text{PCIe}} = \frac{\text{Data}_{\text{PCIe}}}{B_{\text{PCIe}}} 
\label{equ:pci_model_latency}
\end{align}

\begin{equation}
\text{Overall latency} = \text{Latency}_{\text{PCIe}} + \text{Latency}_\text{total}
\end{equation}

\textbf{Resource Usage Model:}
A model for estimating resource usage as a function of the configurable parameters $T_r$, $T_c$ and $T_p$ was developed. 
Eq. (\ref{equ:dsp_usage}) and (\ref{equ:bram_usage}) model resource usage.
%
\begin{align}
&\text{DSP blocks} = \underbrace{(T_r \cdot T_c)}_\text{\# of PEs} ~~\cdot \underbrace{V}_\text{DSPs/MAC unit}
\label{equ:dsp_usage}\\
&\text{BRAM} = WL \left( \underbrace{ T_r \cdot T_p}_\text{buffer $\textbf{A}$} + \underbrace{T_p \cdot T_c}_\text{buffer $\textbf{B}$} + \underbrace{T_r \cdot T_c \cdot (Q+1)}_\text{buffer $\textbf{C}$} \right)
\label{equ:bram_usage}
\end{align}
The factor of $Q$$+$$1$ for buffer C is due to interleaving.


\section{Evaluation} \label{sec:evaluation}
The performance of the tool was evaluated on the Xilinx Virtex UltraScale+ XCVU9P FPGA hosted on the F1 instances on Amazon Web Services (AWS)~\cite{f1_aws}. 
This device has 2586k logic cells, 6840 DSPs and 75.9Mb BRAM. 
The DSPs take $Q$$=$$10$ cycles for an FP32 multiply and $V$$=$$5$ DSPs are used per FP32 MAC unit. 
For INT8 operations, $Q$$=$$1$ and $V$$=$$1$. 

An FP32 GEMM accelerator (Section~\ref{sec:accel_arch}) was developed using Vivado HLS and synthesised using SDAccel 2018.2.
The design was clocked at 250 MHz and the accelerator was configured with $\left<T_r, T_c, T_p\right>$ set to $\left<16, 16, 64\right>$ using the performance model (Section~\ref{sec:perf_model}). 
This was the highest performing design that would route with the current HLS implementation. 
It used 18.8\%, 10.8\%, 8.8\% and 14.1\% of the available DSPs, LUTs, FFs and BRAM respectively.
This design was verified locally on a Xilinx Alveo U250 FPGA. 
Two widely used CNNs, AlexNet~\cite{krizhevsky_imagenet_2012} and ResNet20~\cite{He_2016} were trained on the CIFAR10 dataset, and \textit{Barista} was compared with the CPU \dafinal{and GPU} implementation on Caffe \cite{caffe_2014}.

\begin{figure}[t]
	\centering
	\includegraphics[width=0.95\linewidth, trim=0 10 0 0,clip]{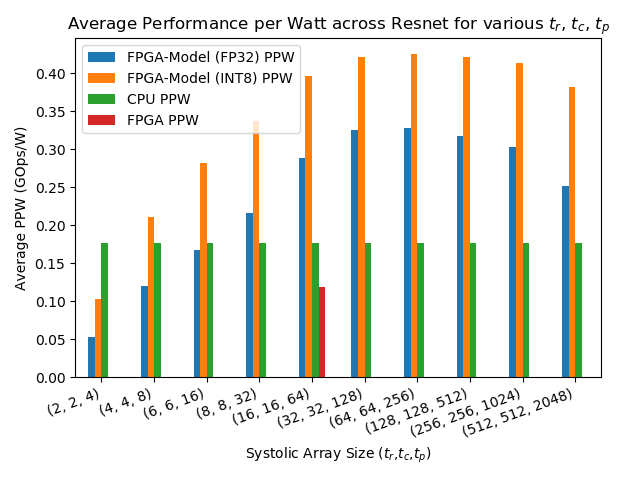}
	\caption{Average PPW across ResNet20 for various $\left<T_r,T_c,T_p\right>$.}
	\label{fig:ppw_comparisons}
\end{figure}

\begin{figure}[t]
	\centering
	\begin{tabular}{c}
	    \subfloat[Implemented]{\includegraphics[width=0.8\linewidth]{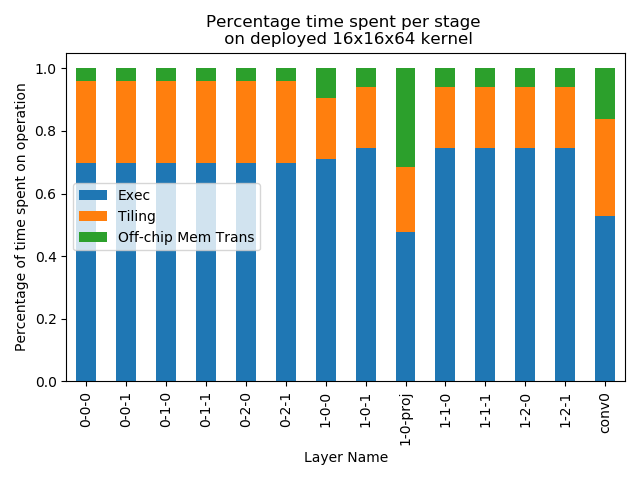} \label{fig:resnet20_ops_perc_true}} \\
	    \subfloat[Model]{\includegraphics[width=0.8\linewidth]{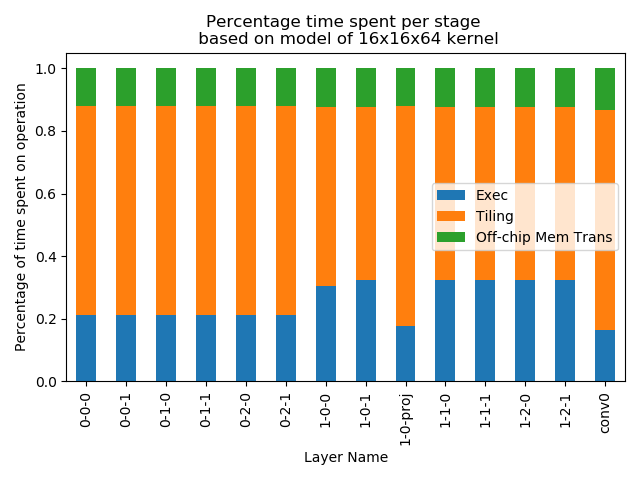} \label{fig:resnet20_ops_perc_model}}
	\end{tabular}
	\caption{Relative time spent on each stage for various ResNet20 layers. Layer names have format (group-residual block-conv).}
	\label{fig:ops_perc}
\end{figure}

Figure \ref{fig:ppw_comparisons} shows the average performance-per-watt (PPW) measured in GOp/s/watt across all CONV layers during the training process (i.e. forward and backward passes) of ResNet20 using \textit{Barista} for the FPGA and CPU. 
\dafinal{For the GPU, the average PPW across all Resnet20 CONV layers was 1.54.}
AWS power profiling showed that the FPGA used 8W of power when running these designs, compared to the CPU's (Intel Xeon E5-2686v4@2.3GHz) 145W and \dafinal{GPU's (NVIDIA GTX 1080Ti) 279W}. 
Profiling was performed using Caffe's internal timers and the design was verified by comparing the FPGA's to the CPU's output.
It is seen that for all sizes of kernel larger than $\left<8,8,32\right>$, both the FP32 (blue bars) and INT8 (orange bars) model predictions outperform the CPU (green bars). 
Additionally, for sizes of kernel larger than $\left<64,64,256\right>$ the performance degrades from performing a large number of zero ops due to tiling (Section~\ref{sec:system_design}) when the kernel size starts to significantly exceed the sizes of the input matrices.
However, the implementation of the $\left<16,16,64\right>$ kernel (red bar) does not outperform the CPU (green bars).

To identify the bottlenecks causing the difference in expected and achieved performance, further profiling was performed using OpenCL. 
Fig.~\ref{fig:resnet20_ops_perc_true} breaks down the relative time spent on each stage of GEMM execution for the implemented kernel using profiled data.
The kernel execution (blue), which includes off-chip to FPGA memory transfers, is seen to be the biggest bottleneck at the moment taking more than 50\% of the time in all CONV layers.
Kernel execution (blue) profiling through Xilinx Vitis' profiler showed that memory bandwidth utilisation for kernel to off-chip memory transfers was in the range of about 10\%. 
Nevertheless, compute unit utilisation rates are at least 70\% indicating the system can be further improved by exploiting memory optimisations.
Fig.~\ref{fig:resnet20_ops_perc_model} shows the same breakdown but now using data from the model for estimates of kernel execution time (blue) and host to off-chip memory transfer time (green). 
Profiled time was used for tiling (orange), which is performed on the CPU. 
The model assumes full utilisation of the DDR4 bandwidth (30Gbps) between off-chip memory and the kernel. 
Fig.~\ref{fig:resnet20_ops_perc_model} demonstrates that with full bandwidth utilisation, the bottleneck is shifted from the FPGA kernel execution (blue) to tiling on the CPU (orange).
Reducing the DDR4 bandwidth assumption to 3Gbps (10\%) in the model predicts a performance close to that achieved by the implemented kernel, supporting the bottleneck analysis from Xilinx's Vitis tool.

\setlength{\tabcolsep}{2pt}
\begin{table}[t]
\centering
\caption{AlexNet predicted best FPGA, CPU and GPU PPW}
\resizebox{\linewidth}{!}{
\begin{tabular}{l c c c c c}
\toprule
\textbf{CONV Layer} & conv1 & conv2 & conv3 & conv4 & conv5 \\ \midrule
$\left<\mathbf{T_r},\mathbf{T_c},\mathbf{T_p}\right>$ & $\left<32,32,74\right>$ & $\left<32,32,64\right>$ & $\left<36,36,64\right>$ & $\left<32,32,64\right>$ & $\left<32,32,64\right>$ \\ 
\textbf{FPGA PPW} & 0.59 & 0.29 & 0.078 & 0.076 & 0.073 \\ 
\textbf{CPU PPW} & 0.35 & 0.24 & 0.089 & 0.13 & 0.11 \\ 
\textbf{GPU PPW} & 0.13 & 0.58 & 0.43 & 0.50 & 0.28 \\ \bottomrule
\end{tabular}
}
\label{tab:alexnet_ppw_per_layer}
\end{table}

A further experiment on tailoring the kernel architecture to the workload was conducted using the model. 
A grid-search was performed across various values of $T_r$,$T_c$ and $T_p$ for designs that are expected to fit on the chosen board based on the memory model.
For ResNet20, the kernel which is predicted to have the highest PPW on average across all layers of the network is $\left<36,36,72\right>$ with a performance of 0.33 GOp/s/W compared to the CPU's 0.18 (+83\%). 
Layer-wise tuning showed that although different-sized kernel performed better on different layers, there is no overall difference in achieved PPW compared to using a single $\left<36,36,72\right>$ kernel for all layers. 
For AlexNet, however, this exploration showed that tailoring the kernel to the layer can provide overall PPW benefits. 
Table~\ref{tab:alexnet_ppw_per_layer} describes the performance of the best kernels per layer and shows that for some layers a CPU performs better than an FPGA for FP32 computations. 
By selectively performing FPGA-based GEMM for conv1 and 2, otherwise using the CPU, the overall achieved PPW is 0.24 compared to the CPUs 0.18 (+33\%) and 0.22 (+10\%) achieved if all layers use one $\left<32,32,64\right>$ kernel. 



\section{Conclusion and Future Work} \label{sec:conc}
Caffe \textit{Barista} enables hardware designers to rapidly prototype novel custom accelerators by seamlessly replacing the provided kernel with one that implements the same interface. 
The model suggests that up to 83\% higher PPW \dafinal{compared to a CPU} can be achieved for a lower absolute power consumption using custom precision arithmetic and/or increasing memory bandwidth utilisation through batching and on-chip tiling. 
From the perspective of a DL researcher, \textit{Barista} allows running any combination of optimisers (\textit{e.g.} SGD, RMSProp, AdaGrad), learning rate schedules and a variety of other training-related parameters or algorithms that are natively supported by or can be implemented in Caffe. 
To the best of our knowledge, \textit{Barista} is the first open-source tool that allows for such versatile and rapid deployment of hardware and algorithms related to the training of CNNs on FPGAs.

\section{Acknowledgements} \label{sec:ack}
The authors would like to acknowledge Huawei for helping fund initial stages of this research. 
Additionally, the authors acknowledge Xilinx for their donation of the Alveo U250 Data Center Acceleration card to collect further results. 
The support of the EPSRC Centre for Doctoral Training in High Performance Embedded and Distributed Systems (HiPEDS, Grant Reference EP/L016796/1) is gratefully acknowledged.

\bibliographystyle{IEEEtran}
\bibliography{references}
\end{document}